# Structure, physical properties, and magnetically tunable topological phases in topological semimetal EuCuBi

Xuhui Wang[1,2,#], Boxuan Li[1,2,#], Liqin Zhou[1,2], Long Chen[1,2], Yulong Wang[1,2], Yaling Yang[1,2], Ying Zhou[1,2], Ke Liao[1,2], Hongming Weng[1,2,3,*], Gang Wang[1,2,3,*]

[1] Beijing National Laboratory for Condensed Matter Physics, Institute of Physics, Chinese Academy of Sciences, Beijing 100190, China

[2] University of Chinese Academy of Sciences, Beijing 100049, China

[3] Songshan Lake Materials Laboratory, Dongguan 523808, China

*Corresponding author. Email: gangwang@iphy.ac.cn; hmweng@iphy.ac.cn.

## ABSTRACT

A single material achieving multiple topological phases can provide potential application for topological spintronics, whereas the candidate materials are very limited. Here, we report the structure, physical properties, and possible emergence of multiple topological phases in the newly discovered, air-stable EuCuBi single crystal. EuCuBi crystallizes in a hexagonal space group $P6_3/mmc$ (No. 194) in ZrBeSi-type structure with an antiferromagnetic (AFM) ground state below $T_N$ = 11.2 K. There is a competition between AFM and ferromagnetic (FM) interactions below $T_N$ revealed by electrical resistivity and magnetic susceptibility measurements. With the increasing magnetic field, EuCuBi evolves from the AFM ground state with a small amount of FM component, going through two possible metamagnetic phases, finally reaches the field-induced FM phase. Based on the first-principles calculations, we demonstrate that the Dirac, Weyl, and possible mirror Chern insulator can be achieved in EuCuBi by tuning the temperature and applying magnetic field, making EuCuBi a promising candidate for exploring multiple topological phases.

## I. INTRODUCTION

The discovery of topological insulators (TIs) featuring dissipationless gapless edge states, has sparked extensive interest in exploring novel topological states [1-4]. Unlike the gapped bulk states in TIs, topological semimetals have quasiparticles caused by nontrivial band crossings in bulk states [5,6]. Among topological semimetals [7,8], Dirac semimetal (DSM) [5,9-13] and Weyl semimetal (WSM) [5,14-16], besides edge states of TIs [17], can create the quasiparticle excitations analogous to relativistic Dirac fermion or Weyl fermion. And triple point semimetal (TPSM) [18-20] with unique surface states [14] and possible topological Lifshitz transitions [21], hosts three-component quasiparticles that do not possess any direct analogue of elementary particle in the quantum field theory [18]. These quasiparticles emerging from various topological phases can be realized and switched in a single topological magnet by manipulating spin configuration. The tunable topological phases can combine the rich physical properties of quasiparticles and provide platform for spintronic applications, especially enhancing the efficiency and robustness of spintronic devices due to their exotic electronic states and topological stability against local perturbations [22-24]. Although some topological magnets have been reported, single material with long-range magnetic order combined with various nontrivial topological electronic structures is highly desired [25-27].

Ternary pnictides with ZrBeSi-type structure having a hexagonal space group $P6_3/mmc$ (No. 194) have attracted much attention owing to their rich topological properties. For instance, KHgSb with honeycomb HgSb layer hosts an hourglass surface state protected by a glide mirror [28,29].

KZnBi was experimentally confirmed to be a three-dimensional DSM with surface superconductivity [28,29]. Moreover, topological phases in a single ternary pnictide with ZrBeSi-type structure can be tuned. BaAgBi, originally reported as a DSM [30-32], can be tuned into a WSM by Eu doping on the Ba site breaking the time reversal symmetry or a TPSM by Cu doping on the Ag site breaking the inversion symmetry [32]. Theoretically, SrAgAs can go through multiple topological phase transitions between DSM, TPSM, and TI by controlling the content of doping Cu atoms on Ag site [18]. However, it is very challenging to experimentally control the occupancy and content of doping atoms for breaking symmetries and inducing new topological phases. In contrast, tuning the symmetry by changing the spin configuration in magnetic topological materials is much easier [25,33,34]. For example, EuAgP can be switched between nearly TPSM and WSM by controlling the directions of magnetization [25]. Below the Néel temperature, EuAgAs in an antiferromagnetic (AFM) state is a topological mirror semimetal or TPSM. Above the Néel temperature, it is in a paramagnetic (PM) state with a pair of Dirac points [26]. In addition, EuAgAs was experimentally found to possess the topological Hall effect caused by the nontrivial spin textures of real space [35-39]. Moreover, the introduction of magnetic atom into topological materials and the enhanced Berry curvature for magnetism result in other novel properties [40], such as the anomalous Hall effect [41-44], anomalous Nernst effect [45-48], magneto-optical effect [49,50], and magnetic spin Hall effect [51-53].

Here we report the structure, physical properties, and possibly emerging topological phases of EuCuBi single crystal. As revealed by the electrical resistivity, magnetic susceptibility, and specific heat capacity measurements, EuCuBi undergoes an AFM transition at $T_N$ = 11.2 K. Below $T_N$, the presence of weak ferromagnetic (FM) component below $\mu_0H_1$ is probably due to the competition of FM and AFM interactions. By using first-principles calculations, we demonstrate the multiple topological states in EuCuBi and the transition among them. Below $T_N$ and $\mu_0H_2$, EuCuBi is in the AFM ground state and the possible magnetic structure is AFM [100] (Néel vector along the *a* axis). For AFM [100], it may be a mirror Chern insulator where all intersections of conduction and valence bands are broken. Above $\mu_0H_3$, spins align ferromagnetically along the direction of applied magnetic field. By shifting the magnetization direction, the magnetic structures of FM [100], FM [110], and FM [001] can be easily obtained, which all host doubly degenerated Weyl points (The spins are aligned along [100] for magnetic structure FM [100]). Under $\mu_0H_2 < \mu_0H < \mu_0H_3$, EuCuBi undergoes two possible metamagnetic transitions. Above $T_N$, EuCuBi is in the PM state as a DSM harbored by six-fold rotation axis. Combining all the obtained results, the magnetic phase diagram of EuCuBi is established.

## Ⅱ. EXPERIMENTAL DETAILS AND METHODS
**Single crystal growth, crystal structure, and chemical composition**

The EuCuBi single crystals were grown by the high-temperature solution method using Bi as flux. The europium ingot (99.9%, Alfa Aesar), copper powder (99.9%, Alfa Aesar), and bismuth granules (99.995%, Alfa Aesar) were mixed in a fritted alumina crucible set (Canfield Crucible Set) [54] in a molar ratio of 2:1:4 and then sealed in an fused silica ampoule under vacuum. The sealed ampoule was heated to 1273 K, kept for 24 h, and then slowly cooled down to 923 K at a rate of 3 K/h. At this temperature, the crystals were separated from the remaining flux by centrifugation. Shiny and hexagonal-shaped crystals were obtained with a size up to 1 mm × 1 mm × 0.5 mm. The crystal structure was determined by single-crystal X-ray diffraction (SCXRD) on a four-circle diffractometer (Rigaku XtaLAB PRO 007HF(Mo)R-DW, HyPix) at 180 K with multilayer mirror graphite-monochromatized Mo $K_\alpha$ radiation ($\lambda$ = 0.71073 Å) operated at 50 kV and 40 mA. The X-ray diffraction (XRD) data of EuCuBi single crystal were measured on a PANalytical X'Pert PRO diffractometer (Cu $K_\alpha$ radiation, $\lambda$ = 1.54178 Å) operated at 40 kV and 40 mA with a graphite monochromator in a reflection mode ($2\theta$ = 10°–100°, step size = 0.017°). The chemical composition was analyzed by a scanning electron microscope (SEM, Hitachi S-4800) equipped with an electron microprobe analyzer for semi-quantitative elemental analysis in energy-dispersive X-ray spectroscopy (EDS) mode.

**Physical property measurement**

Resistivity and magnetoresistance measurements were performed on a physical property measurement system (PPMS) (Quantum Design, 7 T). Contacts for standard four-probe configuration were established by attaching platinum wires using silver paint, resulting in a contact resistance smaller than 5 Ω with the applied current (about 2 mA) parallel to the crystallographic *ab* plane and the magnetic field perpendicular to the *ab* plane [55]. Magnetic susceptibility measurements were carried out with applied magnetic field parallel and perpendicular to the *ab* plane using the zero-field-cooling (ZFC) and field-cooling (FC) protocols. Isothermal magnetization data were collected at 2 K, 4 K, 6 K, 8 K, 12 K, 20 K, and 50 K under the applied magnetic field up to 16 T parallel and perpendicular to the *ab* plane on a high-field PPMS (Quantum Design, 16 T). Specific heat capacity data were collected on a PPMS below 220 K.

**First-principles calculations**

The first-principles calculations were carried out using a plane-wave basis set and projector augmented wave method [56] encoded in the Vienna Ab initio Simulation Package [57,58] for the calculation of electronic band structure. The generalized gradient approximation (GGA) parameterized by Perdew, Burke, and Ernzerhof pseudopotential was used for the exchange-correlation functional [59]. We adopted the GGA+Hubbard-$U$ method where $U$ = 6.0 eV to deal with the strong correlation effects of the Eu-*f* electrons in the magnetic phases [60]. The energy of the plane-wave cutoff was set to 500 eV. The convergence criterion for the total energy was set to be $10^{-7}$ eV. We used a 9 × 9 × 7 *k*-mesh as a sample of the Brillouin zone in the Monkhorst-Pack scheme [60]. All results take into account the spin-orbit coupling (SOC).

## III. RESULTS AND DISCUSSION

### A. Crystal structure and magnetotransport

TABLE I  Crystallographic data and structure refinement for EuCuBi.

| Empirical formula | EuCuBi |
|---|---|
| Formula weight | 424.50 g/mol |
| Space group / Z | $P6_3/mmc$ (No.194) / 2 |
| Unit cell dimensions | $a$ = 4.6099(1) Å  $\alpha$ = 90° |
|  | $b$ = 4.6099(1) Å  $\beta$ = 90° |
|  | $c$ = 8.5208(4) Å  $\gamma$ = 120° |
| Volume / $d_{cal}$ | 156.817(10) Å$^3$ / 8.990 g/cm$^3$ |
| Reflections collected/R(int) | 986 / 0.0789 |
| Data / restraints / parameters | 72 / 0 / 8 |
| Goodness-of-fit on F$^2$ | 1.067 |
| Final R indices [ I>2sigma(I)] | $R_1$ = 0.0280, $\omega R_2$ = 0.0761 |
| $R$ indices (all data) | $R_1$ = 0.0291, $\omega R_2$ = 0.0779 |

| Largest diff. peak and hole | 1.919 and -2.077 e.Å$^{-3}$ |
|---|---|

The crystallographic data for EuCuBi determined by SCXRD is shown in Table I and Table SI. EuCuBi crystallizes in the hexagonal space group $P6_3/mmc$ (No. 194) having the ZrBeSi-type structure with $a$ = 4.6099(1) Å, $b$ = 4.6099(1) Å, $c$ = 8.5208(4) Å, $\alpha = \beta = 90°$, and $\gamma = 120°$. The Cu and Bi atoms are alternately arranged to form a honeycomb layer in AB stacking along the $c$ axis. The Eu atoms are located between the neighboring CuBi layers. The XRD pattern of EuCuBi single crystal is shown in Fig. 1(a), where only the (00$l$) ($l$ = even) diffraction peaks are observed, indicating that the hexagonal-shaped plane is the crystallographic $ab$ plane.

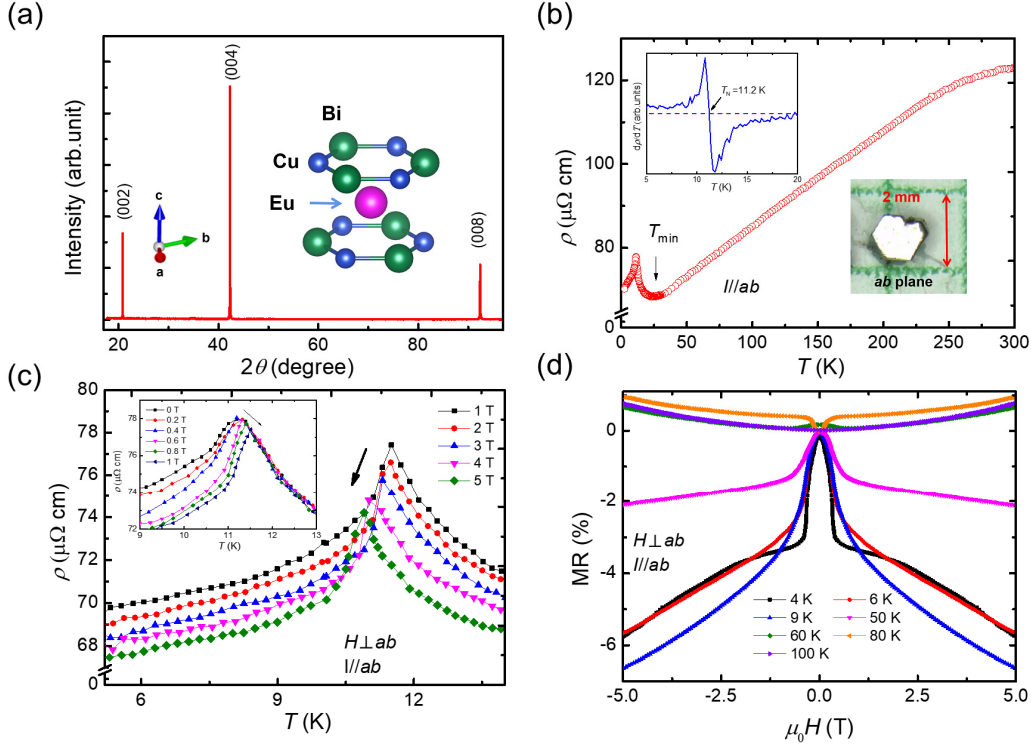

FIG. 1. (a) XRD pattern of EuCuBi single crystal. The inset shows the schematic crystal structure of EuCuBi. (b) Temperature dependence of resistivity for $I // ab$. The upper inset shows the derivative of $\rho$ ($d\rho/dT$). The lower inset shows the optical photograph of EuCuBi single crystal. (c) Temperature-dependent resistivity for $H \perp ab$ (1 T - 5 T and below 1 T in the inset) and $I // ab$. (d) TMR at different temperatures up to 100 K.

The temperature-dependent electrical resistivity of EuCuBi single crystal measured in the temperature range of 2 K-300 K with the current being parallel to the $ab$ plane ($I // ab$) is shown in Fig. 1(b). It has a residual resistivity ratio (*RRR*) ($\rho_{300 K}/\rho_{T\min}$) of about 1.79 (see Fig. 1(b)), close to those of EuAgAs (*RRR* = 1.96) [39], EuAuAs (*RRR* = 2.07) [61], and EuCuAs (*RRR* = 1.35) [62]. The resistivity at 300 K is 122.97 μΩ cm, which is comparable to those of EuAgAs (100 μΩ cm) [39], EuCuAs (110 μΩ cm) [62], and SrAuAs (126 μΩ cm) [63] but lower than that of EuAuAs (430 μΩ cm) [61]. The resistivity gradually decreases with decreasing temperature above $T_{min}$ = 24 K, exhibiting a metallic-like behavior. Below $T_{min}$, the resistivity first increases and then decreases, exhibiting a sharp anomaly at $T_N$ = 11.2 K. The decrease in resistivity below $T_N$ is due to the reduced carrier scattering for ordered Eu$^{2+}$ moments.

By applying magnetic field perpendicular to the $ab$ plane ($H \perp ab$), this anomaly is gradually suppressed with the magnetic field increasing from 1 T to 5 T, exhibiting the low-temperature shifting and intensity decrease (Fig. 1(c)). Such behavior is typical for an AFM transition [64]. Notably, below the magnetic field of 1 T, the anomaly moves towards higher temperature as the magnetic field increases (the inset of Fig. 1(c)), showing a distinct behavior of FM transition [65,66]. Thus, there probably is a competition between FM and AFM interactions [67,68].

The magnetic-field-dependent transverse magnetoresistance (TMR) defined by $[\rho(H) - \rho(0)]/\rho(0)$ × 100% at different temperatures is shown in Fig. 1(d) ($I // ab$, $H \perp ab$). TMR is negative with a cusp-like feature due to the suppression of magnetic fluctuation by increased applied magnetic field below 50 K (well above $T_N$ = 11.2 K) and becomes positive above 50 K. This suggests that the magnetic fluctuation exists well above $T_N$. Under a magnetic field of 5 T, TMR reaches -5.73 % at 4 K, -6.63% at 9 K, and being smaller than 1% at 100 K. TMR at 9 K is smaller than that at 4 K, similar to that of EuAgAs [39], which may be due to the larger magnetic fluctuation close to $T_N$ resulting in more significant magnetic field suppression.

**B. Magnetic properties**

Fig. 2(a) and 2(b) shows the ZFC magnetic susceptibility and its inversion for EuCuBi single crystal with a magnetic field of 0.1 T parallel ($\chi_{//}$) and perpendicular ($\chi_\perp$) to the $ab$ plane at temperature ranging from 2 K to 300 K, respectively. $\chi_{//}$ shows sharper peak at 11.2 K than $\chi_\perp$ at 11 K, suggesting that the magnetic moments align in the $ab$ plane [62,69,70]. The extended Curie-Weiss Law $\chi(T) = C/(T - \Theta_P) + \chi_0$, where $\chi_0$ is the temperature-independent susceptibility, $C$ the Curie constant, and $\Theta_P$ the PM Weiss temperature, is used to fit the magnetic susceptibility at high temperature. The parameters and the effective magnetic moment derived by the fitting are listed in Table II. The estimated effective magnetic moment is 7.71 $\mu_B$ per formula unit for $H // ab$ and 7.77 $\mu_B$ per formula unit for $H \perp ab$, respectively. Both are close to the theoretical value $g\sqrt{S(2S+1)}\mu_B$ = 7.94 $\mu_B$ with $g$ = 2 and $S$ = 7/2 for a free Eu$^{2+}$. The obtained Weiss temperature $\Theta_p$ is negative, -10.99 K and -11.17 K for $H // ab$ and $H \perp ab$, respectively. This indicates the predominance of AFM interaction among Eu$^{2+}$ moments. Besides, there is almost no difference between $\chi_{//}$ and $\chi_\perp$ above 50 K, indicating that the susceptibility of EuCuBi is quite isotropic in the high temperature range.

Figs. 2(c) and (d) show the ZFC and FC curves under certain magnetic field parallel and perpendicular to the $ab$ plane, respectively. Around $T_N$, $\chi_{//}$ increases and then decreases rapidly with decreasing temperature forming a distinct peak associated with the AFM transition under low magnetic field (the inset of Fig. 2(c)). Whereas for $\chi_\perp$, it exhibits a nonmonotonic behavior below $\mu_0H_1$ (Fig. S2), which first decreases and then increases as the temperature drops (the inset of Fig. 2(d)). Such behavior also implies the competition between FM and AFM interactions, similar to that of EuAgAs [39]. Above $\mu_0H_1$, both $\chi_{//}$ and $\chi_\perp$ have magnetic transition that shifts to lower temperature with increasing magnetic field (Fig. 2(c) and (d)), which corresponds to the AMF order. Below $\mu_0H_1$, $d\chi_{//}/dT$ vs. $T$ and $d\chi_\perp/dT$ vs. $T$ both have peak that shifts towards higher temperature with increasing magnetic field in common with FM order (Figs. S2(c) and (d)). This is coincident with resistivity and susceptibility under low magnetic field. The hysteresis loop can be seen in $M(H)$ curves for $H // ab$ and $H \perp ab$, again supporting the existence of FM component (Figs. S2(a) and (b)).

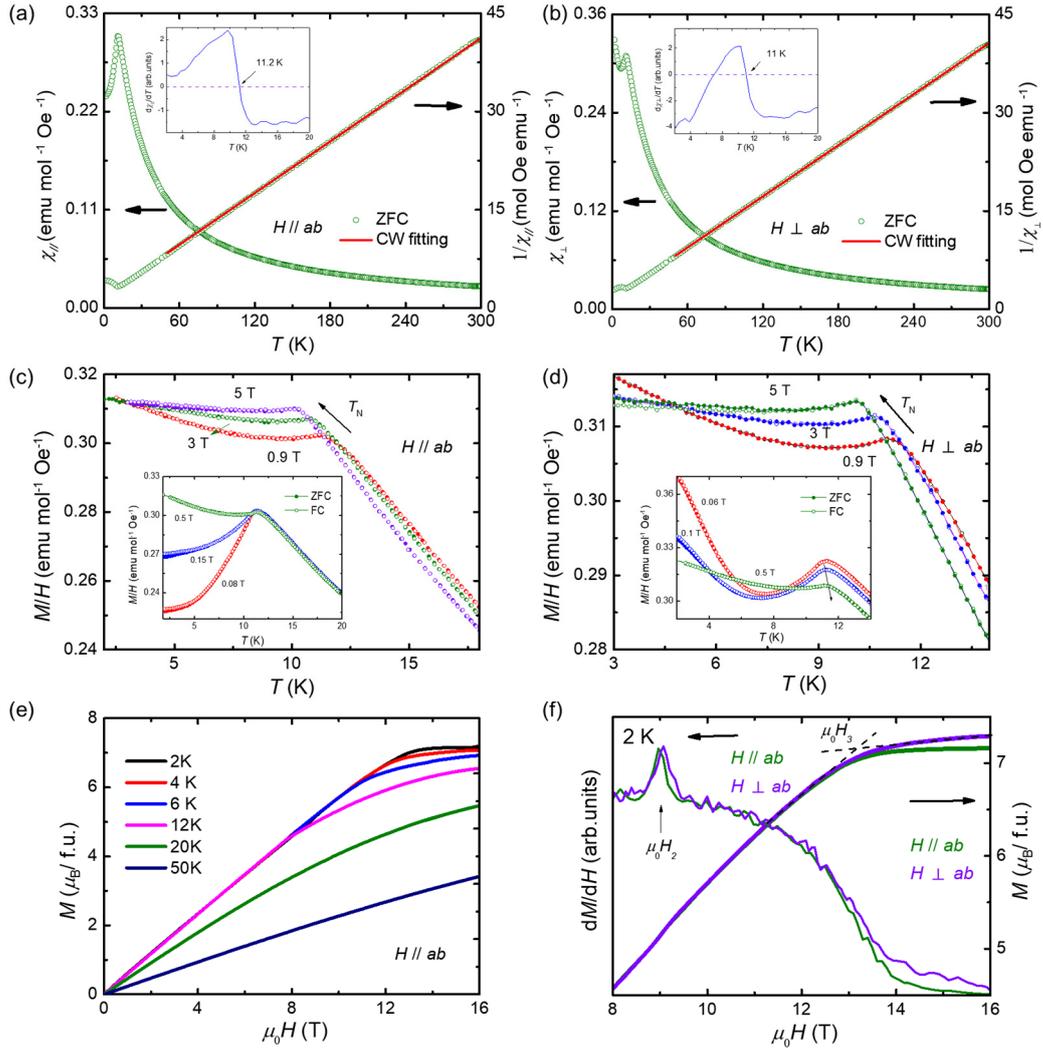

FIG. 2. Temperature dependence of magnetic susceptibility for EuCuBi single crystal for (a) $H // ab$ and (b) $H \perp ab$. The red solid lines show the Curie-Weiss fittings. Magnetic susceptibility vs. $T$ measured in ZFC (filled circle) and FC (hollow circle) protocols for (c) $H // ab$ and (d) $H \perp ab$ (0.9 T < $\mu_0 H_1$). The insets of (c) and (d) show the magnetic susceptibility under magnetic field below 1 T. (e) Magnetic field dependence of isothermal magnetization up to 16 T for EuCuBi single crystal at different temperature for $H // ab$. (f) Isothermal magnetization and the derivative of magnetization ($dM/dH$) for $H // ab$ and $H \perp ab$ at 2 K.

TABLE II. Fitting parameters for the magnetic susceptibility of EuCuBi single crystal at temperature ranging from 50 K to 300 K using the extended Curie-Weiss Law.

| parameter | $H // ab$ | $H \perp ab$ |
| --- | --- | --- |
| $\chi_0$ (×10$^{-4}$ emu/mol Oe) | 3.94 | 4.22 |
| $C$ (emu K/mol Oe) | 7.43 | 7.54 |
| $\Theta_P$ (K) | -10.99 | -11.17 |
| $\mu_{eff}$ ($\mu_B$/f.u.) | 7.71 | 7.77 |

Isothermal magnetization was measured in both directions ($H // ab$ and $H \perp ab$) as shown in Fig. 2(e) and (f). At 2 K, the magnetization increases almost linearly and then gradually saturates to 7.15 $\mu_B$ and 7.29 $\mu_B$ per formula unit for $H // ab$ and $H \perp ab$ respectively, with FM-like spin alignment along the direction of applied magnetic field. Both saturated values are a little bit larger than the theoretical saturated magnetic moment $gS\mu_B = 7\ \mu_B$ for $Eu^{2+}$. Despite small magnetocrystalline anisotropy (consistent with theoretical results below), the easy axis of magnetization should lie in the *ab* plane. At 50 K, the magnetization shows linear magnetic field dependence in the whole measured magnetic field range. Notably, $dM/dH$ vs. $H$ curves have peak around 9 T corresponding to the kink of M-H curves, which exhibit small hysteresis around 9 T for $H\ //\ ab$ and $H \perp ab$ (Figs. S3 (a) and (b)). This suggests that the first metamagnetic (MM1) transition is underway. As the magnetization is clearly not saturated up to $\mu_0H_3$ (see Fig. 2(f)), there may exist the second metamagnetic (MM2) transition below $\mu_0H_3$ (magnetic moments continue to rotate towards the magnetic field).

The isostructural compounds, like EuAgAs [39,71], EuCuAs [62], and EuAuAs [61], all have the A-type AFM order, in which $Eu^{2+}$ is in the FM alignment within the *ab* plane and in the AFM alignment along the *c* axis. Considering the very similar magnetic susceptibility and small magnetic anisotropy, EuCuBi most probably also has the A-type AFM order.

TABLE Ⅲ. Summary of the magnetic properties and inferred magnetic structure from experimental and theoretical results of several materials with ZrBeSi-type structure [39,61,62].

| Materials | Magnetic ground state | AFM & FM competition | $T_N$ (K) | $\Theta_p$ (K) $H\ //\ ab$ | $\Theta_p$ (K) $H \perp ab$ | Magnetic structure (A-type AFM) | Ref. |
|---|---|---|---|---|---|---|---|
| EuCuAs | AFM | | 14 | 19.7 | 18.4 | Spins within *ab* plane | [62] |
| EuAgAs | AFM | √ | 12 | 10.4 | 8.7 | AFM [110] | [39] |
| EuAuAs | AFM | | 6 | 6.7 | 4.1 | AFM [001] | [61] |
| EuCuBi | AFM | √ | 11.2 | -10.99 | -11.17 | AFM [100] | This work |

## C. Heat capacity

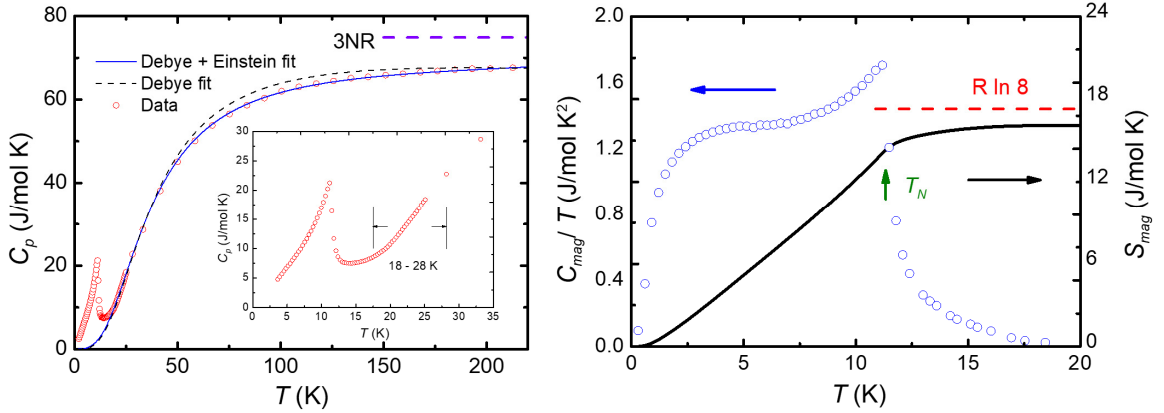

FIG. 3. (a) Temperature dependence of specific heat capacity of EuCuBi single crystal, which is fitted by the Debye model and the sum of Debye and Einstein model. The inset displays the zoomed magnetic transition peak. (b) $C_{mag}/T$ vs. $T$ of EuCuBi single crystal (blue hollow circle) and the calculated magnetic entropy $S_{mag}$ vs. $T$ shown by the solid line.

The specific heat capacity $C_P(T)$ of EuCuBi single crystal was measured under zero field in the temperature range of 2 K–213 K and shown in Fig. 3 (a). A distinct $\lambda$-shape peak is observed around 11.2 K, which is consistent with the $\chi(T)$ and $\rho(T)$. The value of $C_P(T)$ at 213 K is 67.6 J/mol K, which is close to the classical Dulong-Petit limit $C_P = 3NR = 74.83$ J/mol K with $N = 3$ for EuCuBi, where $N$ is the number of atoms per formula unit, $R$ the ideal gas constant. The $C_P(T)$ in the temperature range 2 K–213 K was fitted by the Debye model using the expression as follows:

$$C_D(T) = 9NR\left(\frac{T}{\Theta_D}\right)^3 \int_0^{\Theta_D/T} \frac{x^4 e^x}{(e^x-1)^2} dx \tag{1}$$

$$C_P(T) = \gamma T + C_D \tag{2}$$

Where $\Theta_D$ is the Debye temperature, $\gamma$ the Sommerfeld coefficient, $\gamma T$ the electron specific heat term. As shown by the dashed line in Fig. 3(a), the fitting has obvious deviation in the high temperature region and yields an unphysically negative $\gamma$. A much better fitting is obtained by combining the Debye model with Einstein model using the expression as follows [61,72]:

$$C_E(T) = 3NR\left(\frac{\Theta_E}{T}\right)^2 \frac{e^{\Theta_E/T}}{(e^{\Theta_E/T}-1)^2} \tag{3}$$

$$C_P(T) = \gamma T + (1-b)C_E + bC_D \tag{4}$$

Where $\Theta_E$ is the Einstein temperature, $b$ the weighting factor determining the weight of the Debye model in the lattice specific heat capacity. The parameters obtained from this fitting are as follows: $\gamma = 3.11$ mJ/mol K$^2$, $\Theta_D = 151$ K, $\Theta_E = 400$ K, and $b = 0.92$. The Einstein temperature 400 K is higher than that of EuAuAs ($\Theta_E = 313$ K) and EuMg$_2$Bi$_2$ ($\Theta_E = 305$ K), which may be correlated to the high-frequency optical modes [61,72]. In addition, we also fitted $C_P(T)$ in the temperature range of 18 K-28 K by the low temperature limit $T^3$ term and electron specific heat term of Debye model using the expression as follows:

$$C_P(T) = \gamma T + \beta T^3 \tag{5}$$

Fitting parameters $\gamma = 201.8$ mJ/mol K$^2$, $\beta = 0.84$ mJ/mol K$^4$, and $\Theta_D = 177$ K are derived. $\Theta_D$ is obtained by the expression:

$$\Theta_D = \left(\frac{12\pi^4 NR}{5\beta}\right)^{1/3} \tag{6}$$

The large $\gamma$ is due to the magnon contributions of Eu$^{2+}$. The magnetic part of specific heat capacity $C_{mag}(T)$ is obtained by subtracting the fitting curve obtained using Eq. (4) from the sum of Debye model and Einstein model. The $S_{mag}(T)$ is derived by the expression:

$$S_{mag} = \int_0^T \frac{C_{mag}}{T} dT \tag{7}$$

As shown in Fig. 3(b), $S_{mag}(T)$ tends to saturate and is close to the theoretical value $R\ln(2S+1) = 17.3$ J/mol K for $S = 7/2$ at higher temperature. At $T_N$, $S_{mag}$ achieves a value of 16.09 J/mol K and reaches 85.5 % of $R\ln(8)$ J/mol K. In view of the classical $\lambda$-shape specific heat peak, the corresponding transition should belong to the typical second-order magnetic transition [73-75].

Combining the results presented above, the magnetic phase diagram of EuCuBi is summarized in Fig. 4. The transition temperatures obtained from the magnetic susceptibility are determined by $d\chi/dT$. The critical magnetic fields are determined by $dM/dH$ from the magnetization. More details can be seen in Supporting Information. There are five regions in the magnetic phase diagram: below $T_N$ and $\mu_0H_2$, spins align antiferromagnetically; above $\mu_0H_3$, spins align ferromagnetically along the applied magnetic field direction; when $\mu_0H_2 < \mu_0H < \mu_0H_3$, spins undergo the possible MM1 and MM2 transitions; and in the low-field region above $T_N$, EuCuBi is in the PM state. A weak FM component may exist in the region below $T_N$ and $\mu_0H_1$.

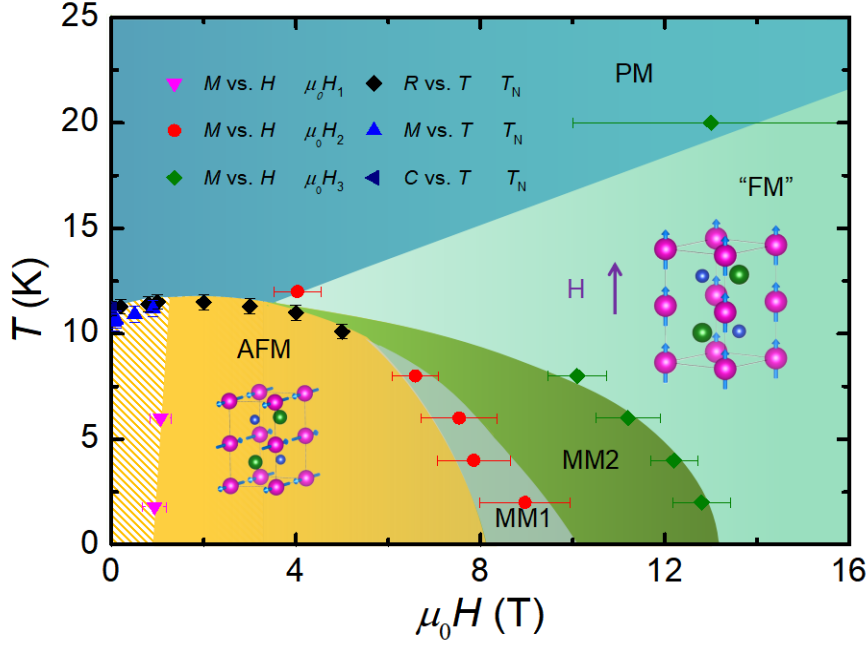

FIG. 4. Magnetic phase diagram for EuCuBi with $H \perp ab$. The possible magnetic phases are shown in the corresponding regions. Different symbols indicate critical temperatures determined by different measurements.

### D. Electronic structure

Within the unit cell of EuCuBi, Eu is at the Wyckoff position of $2a$, Cu at the position of $2d$, and Bi at the position of $2c$. It is noticed that this structure contains a $C_6$ rotation axis and a vertical $\sigma_v$ mirror plane, which protect the crossing points on the high symmetry axis. The degeneracy points near the Fermi energy mainly come from the $s$ orbitals of Cu and Bi, and $d$ orbitals of Eu, whereas the magnetism comes from the $f$ orbitals of Eu.

To figure out the possible magnetic configurations of EuCuBi bleow $T_N$, the total energy of three probable AFM spin configurations, AFM [001], AFM [100], and AFM [110], was calculated. As shown in Fig. 5(c), the total energy of AFM [100] is the lowest, showing the magnetic moments align in the $ab$ plane. But the energy difference between AFM [100] and AFM [001] is very small (33 μeV), also indicating small magnetocrystalline anisotropy. For the self-consistent calculation, AFM [110] is hard to converge, so the convergence criterion for its electronic energy is set to be $10^{-6}$ eV, while all the numerical tolerance for other configurations is set to be $10^{-7}$ eV. We also calculated the total energy of three probable FM configurations: FM [100], FM [110], and FM [001]. Among them, the total energy of FM [001] is the lowest but still higher than that of AFM [100], further indicating AFM state is energetically favorable as the ground state. Above $T_N$, EuCuBi is in the PM state. The spin configurations of FM [001], [100], and [110] can be obtained by applying magnetic field along the [001], [100], and [110] directions till the magnetization reaches saturation at which the spins align along the magnetic field direction.

We then calculated the topological features of different magnetic configurations for EuCuBi. The calculated band structure along the Γ-A path is shown in Fig. 5(f). For PM EuCuBi, the conduction and valence bands intersect at the point D. Since both time reversal and inversion symmetries are preserved in the PM state, all bands are doubly degenerated. The magnetic little co-group of Γ-A path is $6/m'mm$ and the conduction and valence bands belong to the two-dimensional irreducible co-representations DT7 and DT9, respectively. The point D is a fourfold degenerated Dirac point protected by $C_6$ symmetry. Next, we considered the band structure of AFM [001] configuration. The Dirac point is broken for the absence of time reversal symmetry. The doubly degenerated band and the two non-degenerated bands intersect and form two triple degenerated points $T_1$ and $T_2$. The double degenerated band belongs to the irreducible co-representation DT6 of the magnetic little group while the other two bands belong to DT4 and DT5, respectively. For the AFM [100] configuration, the intersections between the conduction and valence bands are all destroyed. Since we find the non-trivial mirror Chern number for the mirror $M_z$, AFM [100] phase of EuCuBi is a possible mirror Chern insulator. In addition, the band structure of FM [001] configuration is also shown in Fig. 5(f). Its magnetic little co-group of Γ-A path is 62'2', in which there are only one-dimensional irreducible representations. The non-degenerated conduction and valence bands intersect at the $W_1$ point, forming a doubly degenerated Weyl point protected by $C_6$ symmetry. Finally, the magnetic little co-group of Γ-A path in FM [100] and FM [110] configurations is $m'm2'$. The conduction bands and the valence bands belong to one-dimensional irreducible representations G4 and G3 respectively, and they intersect at points $W_2$ and $W_3$ forming doubly degenerated Weyl points.

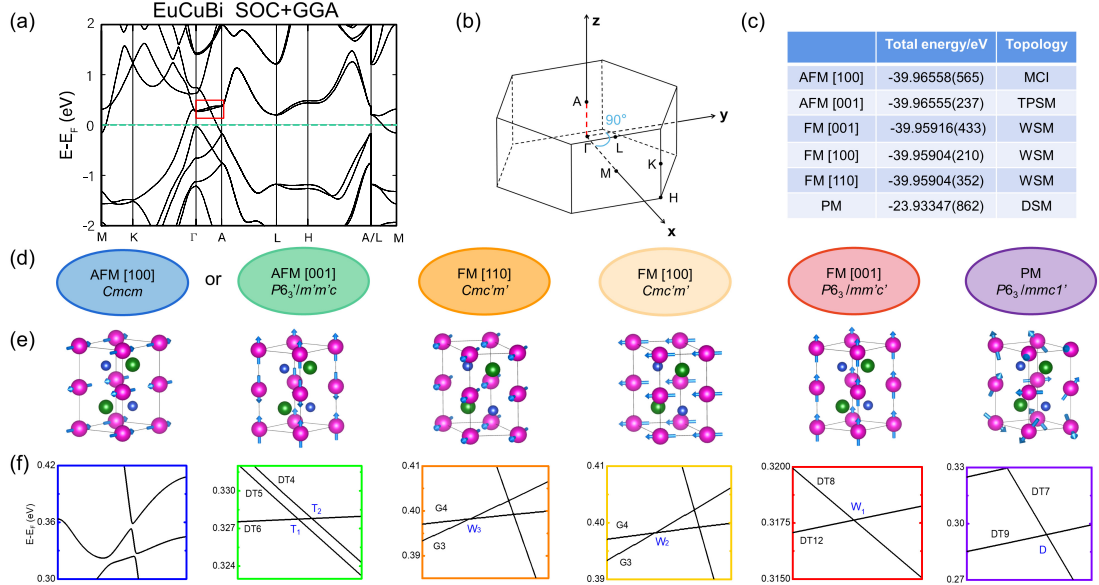

FIG. 5. (a) Calculated band structure of the PM phase with SOC. (b) Bulk Brillouin zone (BZ) of EuCuBi. (c) Total energy and topological features of different magnetic phases. (d) Magnetic space groups of the different accessible phases. (e) Schematic of the spin configurations. (f) Effect of different magnetic structures on the band structure along the Γ-A path.

## IV. CONCLUSIONS

In summary, EuCuBi crystallizes in a hexagonal $P6_3/mmc$ space group with honeycomb layers formed by Bi and Cu atoms, in which the six-fold rotation axis protects the degenerated points on the Γ-A path. An AFM transition at $T_N$ = 11.2 K is confirmed by the resistivity, magnetization, and specific heat capacity measurements. The presence of weak FM component below $T_N$ and $\mu_0H_1$ suggests that there is a competition between FM and AFM interactions. A magnetic phase diagram of EuCuBi is established. Below $T_N$, the symmetries influenced by the long-range magnetic order lead to the emergence of multiple topological states, including TPSM, WSM, and possible mirror Chern insulator. The magnetic phases with different topological states can be effectively obtained by tuning the magnetic field and temperature. Magnetic-tuned topological phase transitions may provide a new perspective on spintronics. The results indicate that EuCuBi should be a promising candidate for revealing the interplay of topology and magnetism and exploring applications of spintronics.

ACKNOWLEDGEMENTS

X. H. Wang and B. X. Li contributed equally to this work. X. H. Wang and G. Wang would like to thank Prof. X. L. Chen of the Institute of Physics, Chinese Academy of Sciences and Dr. Z. N. Guo of the University of Science and Technology Beijing for helpful discussion. This work was partially supported by the National Natural Science Foundation of China (51832010) and the National Key Research and Development Program of China (2018YFE0202600 and 2022YFA14303900).

# Supporting Information

# Structure, physical properties, and magnetically tunable topological phases in topological semimetal EuCuBi

Xuhui Wang[1,2,#], Boxuan Li[1,2,#], Liqin Zhou[1,2], Long Chen[1,2], Yulong Wang[1,2], Yaling Yang[1,2], Ying Zhou[1,2], Ke Liao[1,2], Hongming Weng[1,2,3,*], Gang Wang[1,2,3,*]

[1] Beijing National Laboratory for Condensed Matter Physics, Institute of Physics, Chinese Academy of Sciences, Beijing 100190, China
[2] University of Chinese Academy of Sciences, Beijing 100049, China
[3] Songshan Lake Materials Laboratory, Dongguan 523808, China

**Chemical composition**

TABLE SI. Atomic coordinates and equivalent isotropic displacement parameters for EuCuBi.

| Atom | Wyck. | Site symm. | x/a | y/b | z/c | Occ. | U(eq)(Å$^2$) |
|---|---|---|---|---|---|---|---|
| Bi | 2c | -6m2 | 0.33333 | 0.66667 | 0.75000 | 1.000 | 0.007 |
| Eu | 2a | -3m. | 0.00000 | 1.00000 | 0.50000 | 1.000 | 0.007 |
| Cu | 2d | -6m2 | 0.6667 | 0.33333 | 0.75000 | 1.000 | 0.016 |

The chemical composition of EuCuBi single crystal was analyzed by a scanning electron microscope (SEM, Hitachi S-4800) in energy-dispersive spectroscopy (EDS) mode. The normalized Eu : Cu : Bi ratio is 1.00 (1): 1.00 (1) : 1.01(1).

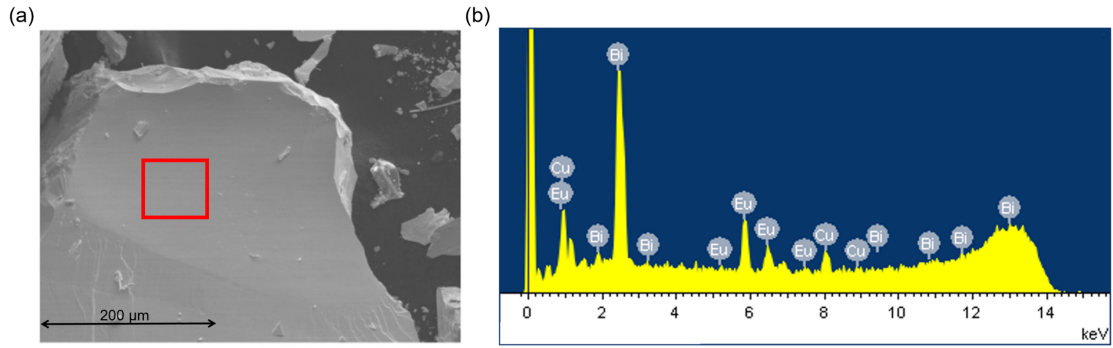

FIG. S1. (a) SEM image and (b) EDS spectrum of EuCuBi.

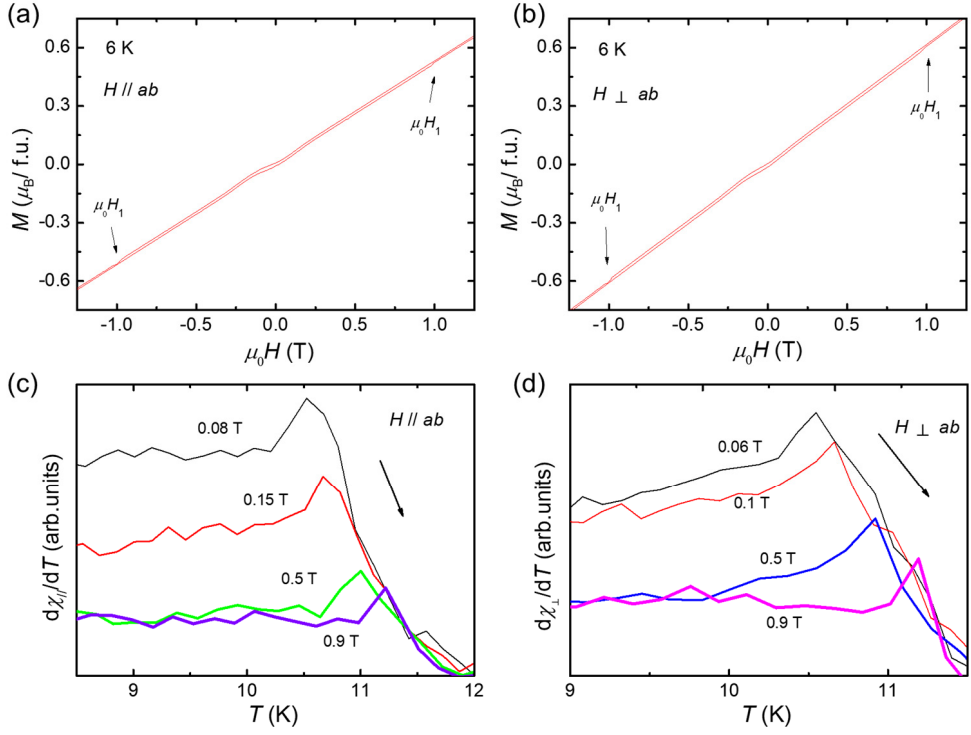

FIG. S2. Isothermal magnetization of EuCuBi single crystal with (a) $H // ab$ and (b) $H \perp ab$ at 6 K. The derivative of magnetic susceptibility for (c) $H // ab$ and (d) $H \perp ab$ below 1 T.

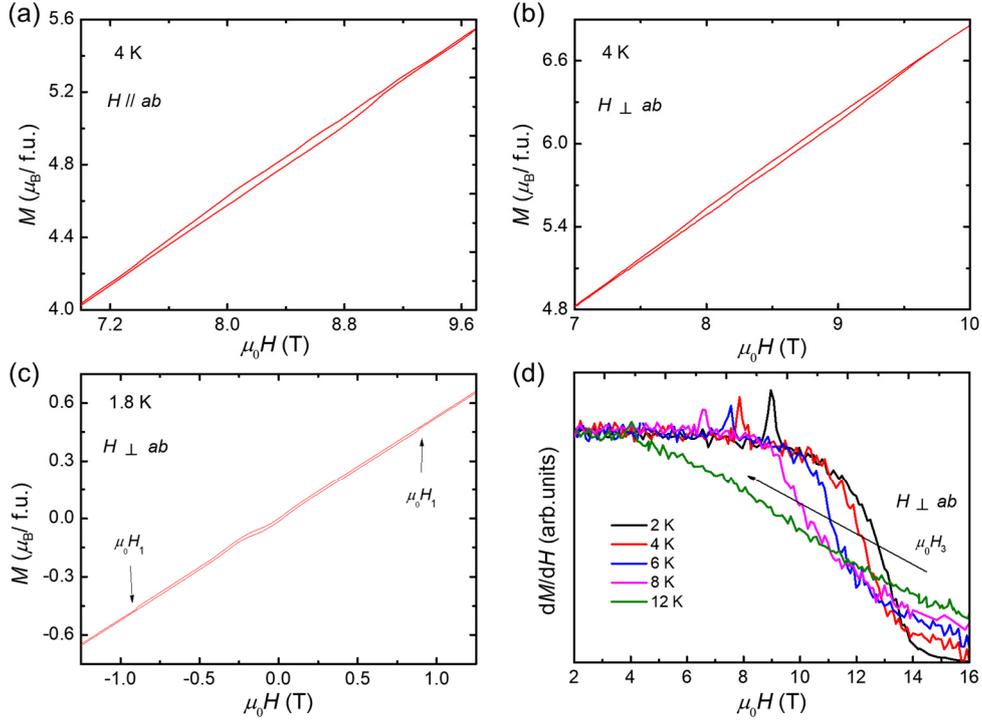

FIG. S3. Isothermal magnetization of EuCuBi single crystal with (a) $H // ab$ and (b) $H \perp ab$ at 4 K around 9 T and (c) with $H \perp ab$ at 1.8 K. (d) The derivative of magnetization at different temperatures for $H \perp ab$.

# Calculated topological properties of EuCuBi

TABLE SII. Character table of little group on the Γ-A path. Here k = (0, 0, ω).

|  | Little group |  | $E$ | $C_2$ | $C_3$ | $C_6$ | $\sigma_v$ |
|---|---|---|---|---|---|---|---|
| PM | $C_{6v}$ (6mm) | DT7 | 2 | 0 | -2 | 0 | 0 |
|  |  | DT9 | 2 | 0 | 1 | $\sqrt{3}e^{i\pi\omega}$ | 0 |
| AFM [001] | $C_{3v}$ (3m) | DT4 | 1 |  | -1 |  | $e^{i\pi(-1+2\omega)/2}$ |
|  |  | DT5 | 1 |  | -1 |  | $e^{i\pi(1+2\omega)/2}$ |
|  |  | DT6 | 2 |  | 1 |  | 0 |
| AFM [100] | $C_{2v}$ (mm2) | DT5 | 2 | 0 |  |  | 0 |
| FM [001] | $C_6$ (6) | DT8 | 1 | $e^{i\pi(1+2\omega)/2}$ | -1 | $e^{i\pi(\pm 1+2\omega)/2}$ |  |
|  |  | DT12 | 1 | $e^{i\pi(1+2\omega)/2}$ | $\pm e^{i\pi/3}$ | $e^{i\pi(\pm 5+2\omega)/2}$ |  |
| FM [100] | $C_s$ (m) | (DT) $\bar{B}_3$[G4] | 1 |  |  |  | -i |
|  |  | (DT) $\bar{B}_4$[G3] | 1 |  |  |  | i |
| FM [110] | $C_s$ (m) | (DT) $\bar{B}_3$[G4] | 1 |  |  |  | -i |
|  |  | (DT) $\bar{B}_4$[G3] | 1 |  |  |  | i |

TABLE SIII. Magnetic space groups (MSG), unitary subgroups, little croup, magnetic little co-group, and topological features of different spin configurations.

|  | MSG | Unitary subgroup | Little group | Magnetic little co-group | Topology |
|---|---|---|---|---|---|
| PM | $P6_3/mmc1'$ (194.264) | 194 | $C_{6v}$ (6mm) | 6/m'mm | DSM |
| AFM [001] | $P6_3'/m'm'c$ (194.268) | 163 | $C_{3v}$ (3m) | -6'm2' | TPSM |
| AFM [100] | $Cmcm$ (63.457) | 63 | $C_{2v}$ (mm2) | mm2 | TCI |
| FM [001] | $P6_3/mm'c'$ (194.270) | 176 | $C_6$ (6) | 62'2' | WSM |
| FM [100] | $Cmc'm'$ (63.463) | 12 | $C_s$ (m) | m'm2' | WSM |
| FM [110] | $Cmc'm'$ (63.463) | 12 | $C_s$ (m) | m'm2' | WSM |